\documentclass[twocolumn,showpacs,preprintnumbers,amsmath,amssymb,pre]{revtex4}

\usepackage{graphicx}
\usepackage{dcolumn}
\usepackage{bm}

\begin{document}

\title{Discrete light localization in one dimensional nonlinear lattices with arbitrary non locality}

\author{Andrea Fratalocchi}
\email{frataloc@uniroma3.it}
\author{Gaetano Assanto}
 \email{assanto@uniroma3.it}
\affiliation{
NooEL - Nonlinear Optics and OptoElectronics Labs.\\
INFM-CNISM and Dept. of Electronic Engineering, University Roma Tre,
Via della Vasca Navale 84, 00146, Rome, Italy
}

\date{\today}

\begin{abstract}
We model discrete spatial solitons in a periodic nonlinear medium encompassing any degree of transverse non locality. Making a convenient reference to a widely used material -nematic liquid crystals-, we derive a new form of the discrete nonlinear Schr\"odinger equation and find a novel family of discrete solitons. Such self-localized solutions in optical lattices can exist with an arbitrary degree of imprinted chirp and a have breathing character. We verify numerically that both local and non local discrete light propagation and solitons can be observed in liquid crystalline arrays.
\end{abstract}

\pacs{42.65.Jx, 42.65.Tg, 42.70.Dm}

\maketitle

\section{Introduction}
Optical energy localization in lattices has become an important branch of contemporary nonlinear science, due to a wealth of basic physics and potentials for light switching and logics \cite{rev_wa_CristA,rev_wa_Kevre,rev_wa_FleisA} (and references therein). Attention has been recently devoted to light propagation in the frame of \emph{tunable discreteness} -i.e. in lattices with an adjustable period, index contrast, nonlinearity. Examples to this extent are waveguide arrays in photorefractives \cite{rev_wa_Crist}, and droplet arrays of a Bose Einstein condensate \cite{wa_BEC_Tromb,wa_BEC_AbdulA,wa_BEC_Ander,wa_BEC_KartaA}. In nematic liquid crystals (NLC), materials encompassing a large non resonant reorientational response, spectrally extended transparency, strong birefringence and mature technology \cite{NLCK}, excitation as well as switching/steering of discrete solitons has been reported in voltage-tunable geometries \cite{wa_NLC_Frat1A,wa_NLC_fratSA,wa_NLC_FratBA}. Discrete solitons in NLC result from the interplay between evanescent coupling (owing to discreteness), molecular nonlinearity (leading to progressive mismatch as the extraordinary index increases) and non locality (owing to intermolecular elastic forces). The latter aspect, despite its role in several systems \cite{non_PHC_Minga,non_BEC_PerezA,non_PHO_MamaeA,non_BangSA,non_LAT_GaidiA,non_LAT_KartaA,non_DIS_GaidiA,non_DIS_Fedd} including NLC \cite{non_NLC_ContT,non_NLC_ContB}, has only been discussed in the framework of 1D discrete lattices with reference to first order contributions \cite{non_LAT_KartaA} and long range dispersive interactions \cite{non_LAT_GaidiA,non_DIS_GaidiA,non_DIS_Fedd}, a general description of discrete solitons in the presence of a transverse nonlinear non locality being still lacking.\\
In this paper, for the first time and with explicit reference to a physical system of interest -i. e., nematic liquid crystals- we model discrete light localization in media with an arbitrary degree of non locality, elucidating the interplay between non locality and nonlinearity on soliton dynamics. Using coupled mode theory to derive the governing equations (CMT), \cite{dis_Chris} we demonstrate the existence of a new family of chirp-imprinted discrete breathers which could not be sustained by a purely local response. Finally, we verify the theoretical predictions by numerical experiments with a standard NLC. The paper is organized as follows. Section \ref{tapp} introduces a model of liquid crystals and carries out an original reduction to a non integrable discrete nonlinear non local Schr\"odinger equation, outlining the novelties with respect to previous studies dealing with non locality. In Sec. \ref{res} we apply a variational approach using a convenient soliton ansatz and derive the differential equations for the evolution of soliton parameters in propagation. We highlight the impact of non locality on soliton generation and demonstrate a novel family of discrete chirped solitary waves, never reported before. Finally, in Sec. \ref{nres}, we perform a full numerical simulation of the actual liquid crystalline system and demonstrate the excellent agreement with our analytical predictions. We conclude by emphasizing how the examined NLC-lattice offers the rare possibility to observe both local and non local light propagation in one and the same system.
\begin{figure}
\includegraphics{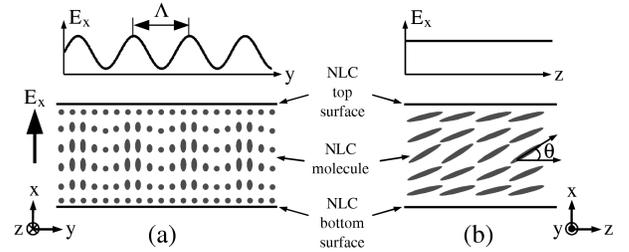}
\caption{\label{nlc_sample}
  Example of a discrete optical lattice in a cell of thickness $d$ filled with undoped nematic liquid crystals: (a) front view, (b) side view. A periodically varying electric field $E_x$ is applied trough an electrode-array across $x$ and alters the mean molecular angular orientation $\theta(x,y)$, inducing an index modulation with the same period $\Lambda$. The top graphs sketch the distribution $E_x$ versus y (left) and versus propagation z (right), respectively.
}
\end{figure}
\section{Theoretical approach}
\label{tapp}
We consider light propagation in a thin film planar waveguide of nematic liquid crystals, subject to a periodic transverse modulation along $y$ and across $x$ (Fig. \ref{nlc_sample}). NLC consist of rod-like molecules which, electrically polarized across $x$, react to and reorient towards the field vector in order to minimize the free energy \cite{NLCK}. Under "planar" anchoring conditions at top and bottom interfaces of the cell, the mean angular orientation of the NLC molecules (i.e., the molecular $director$) is conveniently described by their angle $\theta$ with the axis $z$ in the plane ($x,z$), as sketched in Fig. \ref{nlc_sample}. This identifies the NLC optic axis with respect to the propagation wavevector of a light beam injected in the cell. If $n_a^2=n_{\parallel}^2-n_{\perp}^2$ is the NLC optical birefringence (with $n_{\parallel}$ and $n_{\perp}$ along or orthogonal to the $director$, respectively), an electric field (static or low frequency) applied across $x$, constant in $z$ and periodic along $y$ (Fig. \ref{nlc_sample}), can reorient the director and determine a one-dimensional optical lattice with index modulation $n^2(x,y)=n_{\perp}^{2}+n_{a}^2\sin^2\theta(x,y)$ for $e$-polarized light. We assume an applied electric field $E_x(x,y)=E_0\cdot\sqrt{1+\epsilon F(y)}$, with a zero mean-value $F(y)=F(y+\Lambda)$ and an arbitrary $\epsilon<1$. In actual experiments, $E_x$ is determined by the bias $V(x,y)\propto x(1+V(y))$, with $V(y)=V(y+\Lambda)$ applied through an array of parallel finger-electrodes \cite{wa_NLC_Frat1A}.\\ 
In the framework of the elastic continuum theory \cite{NLCK}, the director distribution $\theta_0(x,y)$ at rest -i.e., with no injected light - can be obtained by minimizing the NLC energy functional
\begin{equation}
\label{dir}
K\nabla_{xy}^{2}\theta_0+\frac{\Delta\epsilon_{RF}\lvert E_{x}\rvert^{2}}{2}\sin(2\theta_0)=0
\end{equation}
$K$ being the NLC elastic constant (single constant approximation \cite{NLCK}) and $\Delta\epsilon_{RF}$ the (low frequency) anisotropy. When an $e$-polarized optical beam of slowly-varying envelope A propagates in the medium, Eq. (\ref{dir}) modifies into:
\begin{equation}
\label{dirc}
K\nabla_{xy}^{2}\theta+\bigg(\frac{\Delta\epsilon_{RF}\lvert E_{x}\rvert^{2}}{2}+
\frac{n_a^2\epsilon_{0}\lvert A\rvert^{2}}{4}\bigg)\sin(2\theta)=0
\end{equation}
The overall director distribution can be written as $\theta(x,y)=\theta_0(x,y)(1+\psi(x,y))$, with $\theta_0(x,y)\psi(x,y)$ the nonlinear optical contribution. In usual experiments $\theta_0$ is small ($\leq 0.4$), hence first order approximations are justified \cite{wa_NLC_Frat1A}. The non locality (linked to the $\nabla$ operator in Eqs. (\ref{dir})-(\ref{dirc})) has a different impact along $x$ and $y$, respectively, due to the strong asymmetry of the problem (Fig. 1). A bell-shaped beam with $x$-waist comparable with the cell thickness $d$ does not experience a non local response along $x$, owing to the planar anchoring with $\theta_0\approx 0$ in $x=\pm d/2$. Conversely, along $y$ no anchoring is present and the index perturbation is free to widen. After substituting $\theta$ into Eqs. (\ref{dir})-(\ref{dirc}), assuming the response to be weakly nonlinear $(\psi\ll 1)$ and local in $x$ $(\partial_x^2(\theta_0\psi)\approx 0)$, we obtain:
\begin{equation}
\label{psic}
\frac{\partial^2\psi}{\partial y^2}+\frac{2}{\theta_0}\frac{\partial\theta_0}{\partial y}\cdot\frac{\partial\psi}{\partial y}-\frac{4\Delta\epsilon_{RF}\lvert E_{x}\rvert^{2}\theta_0^2}{3K}\psi+\frac{n_a^2\epsilon_{0}\lvert A\rvert^{2}}{2K}=0
\end{equation}
Eq. (\ref{psic}) models the all-optical response of the NLC lattice. It can be cast in the integral form $\psi=\iint G(\zeta-x,\eta-y)\lvert A(\zeta,\eta)\rvert^2 d\zeta d\eta$, with $G(x,y)$ the Green function. For a sufficient bias to induce an array of single-mode channel waveguides with nearest-neighbor coupling, CMT (in the tight-binding approximation) yields the $z$ evolution of each eigenmode:
\begin{align}
\label{cmt}
i\frac{\partial Q_n}{\partial z}+C(Q_{n+1}+&Q_{n-1})
\nonumber\\
&+Q_n\sum_m\Gamma_ {m,n}\lvert Q_{m}\rvert^2=0
\end{align}  
with $A=\sum_n{Q_n(z)f_Q(x,y)\exp(-i\beta z)}$, $f_Q$ and $\beta$ the modal eigenfunction and eigenvalue, respectively, $\lvert Q_n\rvert^2$ the mode-power in the $n$th-channel, $C$ the coupling strength and:
\begin{align}
\label{gamma}
 \Gamma_{m,n}&=\frac{\omega\epsilon_0}{4}\int\lvert f_Q(x,y-n\Lambda)\rvert^2 n_{2,\psi}(x,y)
\nonumber\\
&\times G(\zeta-x,\eta-y)\lvert f_Q(\zeta,\eta-m\Lambda)\rvert^2\mathrm{dxdyd\zeta d\eta}
\end{align}
the nonlinear overlap integral, with  $n_{2,\psi}(x,y)=\frac{n_a^2\sin(2\theta_0)\theta_0}{2n(\theta_0)}$. Factorizing the Green function as $G(x,y)=G(x)G_p(y)G_e(y)$, with $G_p(y)=G_p(y+\Lambda)$ and the envelope $G_e(y)$ wider than the guided mode, the integral (\ref{gamma}) becomes $\Gamma_{m+n}=G_e[(m+n)\Lambda]\Upsilon$, with:
\begin{align}
\label{upsi}
\Upsilon=\frac{\omega\epsilon_0}{4}\int\lvert &f_Q(x,y)\rvert^2 n_{2,\psi}(x,y)G(\zeta-x)
\nonumber\\
&\times G_p(\eta-y)\lvert f_Q(\zeta,\eta)\rvert^2\mathrm{dxdyd\zeta d\eta}
\end{align}
To proceed with the analysis, we need to calculate the Green function from Eqs. (\ref{dir}) and (\ref{psic}). By setting $(x,y)=(Xd,Y\Lambda)$, $\theta_0=\theta_r\vartheta(X,Y)$, $\vartheta(0,0)=1$, $\psi=\frac{1}{k_0^2n_a^2\Lambda^2}\phi(X,Y)$, $A^2=\frac{8}{\sqrt 3}\frac{\sqrt{\Delta\epsilon_{RF}K}E_0}{\epsilon_0k_0^2n_a^4\theta_r^2\Lambda^3}a^2(X,Y)$, $\phi=\iint\frac{g(\zeta-X,\eta-Y)}{\vartheta(\zeta-X,\eta-Y)}\lvert a(\zeta,\eta)\rvert^2\mathrm{d\zeta d\eta}$ (scaling $g$ to $\vartheta$ prevents first order derivatives to appear in Eq. (\ref{greenc})), $F=\frac{4\theta_r^2}{3\alpha}f(Y)$, $\alpha=\frac{\Lambda^2}{R_c^2}$, $R_c^2=\frac{3K}{4\Delta\epsilon_{RF}E_0^2\theta_r^2}$, $\beta=\frac{\Lambda^2}{d^2}$ and $\gamma(\epsilon)=(\alpha+\epsilon \frac{4\theta_r^2f}{3})\vartheta^2+\frac{\partial^2\vartheta}{\partial Y^2\vartheta}$, equation (\ref{psic}) with (\ref{dir}) can be cast in the dimensionless form:
\begin{align}
\label{adim}
\frac{\partial^2\vartheta}{\partial Y^2}+\beta\frac{\partial^2\vartheta}{\partial X^2}+(\frac{3}{4\theta_r^2}\alpha+\epsilon f)\vartheta &=0\\
\label{greenc}
\frac{\partial^{2}g}{\partial Y^2}-\gamma(\epsilon)g+2\sqrt\alpha u_0&=0
\end{align}
In order to solve Eq. (\ref{adim}) above, we separate the variables X and Y by letting $\frac{3\alpha}{4\theta_r^2}=\upsilon_x+\upsilon_y$ and obtain a simple form of Hill's equation in Y:
\begin{equation}
\label{adim1}
\frac{d^2\vartheta}{dY^2}+(\upsilon_y+\epsilon f)\vartheta=0
\end{equation}
with $\vartheta(X,Y)=\vartheta(X)\vartheta(Y)$ and $\vartheta(X)\propto\sin(\pi X+\frac{\pi}{2})$ corresponding to a harmonic oscillator across $X$. According to Floquet theory, the periodic solutions $\vartheta(Y)=\vartheta(Y+1)$ are located on a transition curve \cite{NAYP}. We adopt the perturbative method of strained parameters \cite{NAY,NAYP}, performing the following expansion:
\begin{align}
\label{adim10}
\vartheta&=\sum_m\epsilon^m \vartheta_m(Y)
\nonumber\\
\upsilon_y&=\sum_m\epsilon^m \upsilon_{y0}
\end{align}
with $\vartheta_m(Y)=\vartheta_m(Y+1)$. By collecting and equating terms of the same order in $\epsilon$, we find: 
\begin{align}
  \label{sp0}
  \frac{\partial \vartheta_0(Y)^2}{\partial Y^2}&=0 & O(1)\\
  \frac{\partial \vartheta_1(Y)^2}{\partial Y^2}&=-\upsilon_{y1}\vartheta_0-f\vartheta_0 & O(\epsilon)\\
  \label{sp1}
  \frac{\partial \vartheta_2(Y)^2}{\partial Y^2}&=-\upsilon_{y2}\vartheta_0-\upsilon_{y1}\vartheta_1-f\vartheta_1 & O(\epsilon^2)
\end{align}
By expanding $f(Y)$ in a Fourier basis $f=\sum_m \chi_m\exp(-i2\pi mY)$, we obtain the solution to Eqs. (\ref{sp0})-(\ref{sp1}):
\begin{align}
  \label{adim11}
  \vartheta=1&+\sum_m\epsilon^m \vartheta_m(Y)=1+
  \nonumber\\
  &\epsilon \sum_m\frac{\chi_m}{4\pi^2m^2}\exp(-i2\pi mY)+O(\epsilon^2)
\end{align}
\begin{equation}
  \upsilon_y=-\epsilon^2\int\vartheta_1(Y)f(Y)\mathrm{dY}+O(\epsilon^3)
\end{equation}
Substituting Eq. (\ref{adim11}) into Eq. (\ref{greenc}), for $\vartheta(X)\approx\mathrm{constant}$ within the effective modal width, we get:
\begin{equation}
\label{greenc1}
\frac{\partial^{2}g}{\partial Y^2}-\big[\alpha+\sum_m\epsilon^m h_m(Y)\big]g+2\sqrt\alpha u_0=0
\end{equation}
with $g(X,Y)=u_0(X)g(Y)$, $h_1=f(Y)\big[\frac{4\theta_r^2}{3}-\frac{1}{\vartheta(Y)}\big]+2\alpha\vartheta_{1}(Y)$ and a periodic $h_m(Y)=h_m(Y+1)$ the expression of which stems from Eqs. (\ref{sp0})-(\ref{adim11}). For $u_0=0$ Eq. (\ref{greenc1}) is Hill's equation; otherwise for $u_0\ne 0$ its solutions can be found through the expansion:
\begin{equation}
  \label{gfexp0}
  g(Y)=\sum_m\epsilon^m g_m(Y)
\end{equation}
Substituting and equating terms of the same order in $\epsilon$, we have:
\begin{align}
  \label{gfexp1}
  &\frac{\partial g_0^2}{\partial Y^2}-\alpha g_0+2\sqrt\alpha u_0=0 & O(1)\\
  &\frac{\partial g_1^2}{\partial Y^2}-\alpha g_1=h_1 g_0 & O(\epsilon)\\
  \label{gfexp2}
  &\frac{\partial g_2^2}{\partial Y^2}-\alpha g_2=h_1 g_1+h_2 g_0 & O(\epsilon^2)
\end{align}
At each order the solution can be factorized in the form $g(Y)=g_e(Y)g_{pn}(Y)$, i.e. an envelope $g_e(Y)$ modulated by the periodic function $g_{pn}(Y)=g_{pn}(Y+1)$, with:
\begin{widetext}
  \begin{align}
    \label{gfexpf}
    &g_e(Y)=\exp(-\sqrt\alpha\lvert Y\rvert)
    \nonumber\\
    &g_{p0}=1
    \nonumber\\
    &g_{p1}(Y)=-\sum_m \frac{h_{1m}}{4m(\alpha+m^2\pi^2)}\bigg[m+3m\exp(-i2\pi mY)+\frac{2\mathrm{sign(Y)}}{\pi\sqrt\alpha}\exp(-i\pi mY)\sin(m\pi Y) \bigg]
  \end{align}
\end{widetext}
Therefore, the generic solution $g(Y)$ can be written in the form $g_n(Y)=g_e(Y)g_p(Y)$, with a peaked envelope $g_p(Y)$ and a periodic modulation $g_p(Y)=\sum_n\epsilon^n g_{pn}$.
\begin{figure}
\includegraphics{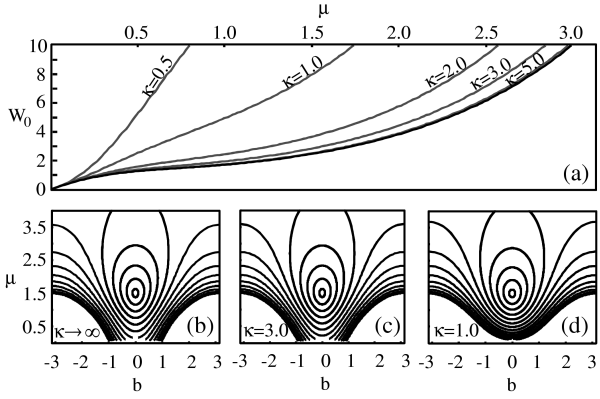}
\caption{\label{power}
(a) Power $W_0$ of discrete soliton versus size $\mu$ for increasing $\kappa$ (decreasing non locality) up to the local Kerr-case (thick line), (b)-(d) Phase planes of soliton chirp $b$ and width $\mu$ for various $\kappa$ in the case $\mu=1.5$.
}
\end{figure}
Now, after introducing the (dimensionless) fields $q_n=Q_n\sqrt{\frac{\Upsilon\coth(\kappa/2)}{2C}}\exp(-i2\xi)$ with $\xi=Cz$ and the degree of (non) locality $\kappa=\sqrt\alpha$, by substituting Eqs. (\ref{gfexpf}) in Eq. (\ref{cmt}) we finally obtain the discrete non local nonlinear Schr\"odinger equation:
\begin{align}
\label{cmtf}
i\frac{\partial q_n}{\partial \xi}&+(q_{n+1}+q_{n-1}-2q_n)
\nonumber\\
&+2q_n\sum_m\frac{\exp(-\kappa\lvert m+n\rvert)}{\coth\frac{\kappa}{2}}\lvert q_{m}\rvert^2=0
\end{align}
Equation (\ref{cmtf}) is a discrete version of the non local nonlinear Schr\"odinger equation (NNLS) \cite{non_BangSA} and reduces to the discrete nonlinear Schr\"odinger equation (DNLS) \cite{wa_Malom} as $\kappa\rightarrow\infty$. At variance with previous work dealing with non locality in dispersion (a linear property of the medium) \cite{non_LAT_GaidiA,non_DIS_GaidiA,non_DIS_Fedd}, Eqs. (\ref{cmtf}) addresses a \emph{nonlinear} feature (non locality in the nonlinear response) and, thereby, it is expected to possess a radically distinct dynamics with respect to its linear counterpart (see Sec. \ref{res}) and to non local models in the frame of nonlinear photonic crystals (\cite{non_PHC_Minga} and references therein). The lack of propagation terms in Eqs. (8)-(9) of \cite{non_PHC_Minga}, for instance, does not allow to study the system evolution (as investigated hereby). Hence, equation (\ref{cmtf}) can be regarded as a novel general model of discrete, dispersive, nonlinear non 
local media.

\section{Analysis of the discrete model}
\label{res}
\begin{figure}
\includegraphics{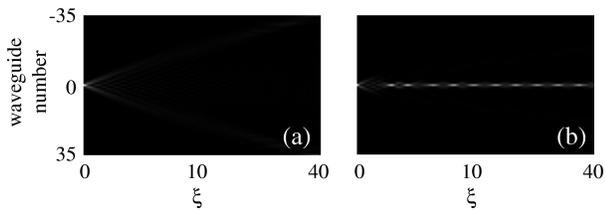}
\caption{\label{sim1}
(a) Discrete soliton breaking in a quasi-local regime $(\kappa=3.0)$ and (b) breather-like propagation in the non local regime $(\kappa=1.0)$ for $(\mu=1.5)$ and an imprinted chirp $b=3$.
}
\end{figure}

Equation (\ref{cmtf}) can be also derived using the variational principle and the following Lagrangian $L$: 
\begin{align}
\label{lag}
L&=\sum_n\frac{i}{2}(q_n^*\frac{\partial q_n}{\partial\xi}-q_n\frac{\partial q_n^*}{\partial\xi})
\nonumber\\
&-\lvert q_{n+1}-q_{n}\rvert^2+\sum_m\frac{\exp(-\kappa\lvert m+n\rvert)}{\coth\frac{\kappa}{2}}\lvert q_{m}\rvert^2\lvert q_n\rvert^2
 \end{align}
We adopt the peaked ansatz $q_n(\xi)=A_q(\xi)\exp[i\varphi(\xi)+ib(\xi)\lvert n-n_0\rvert-\mu(\xi)\lvert n-n_0\rvert]$ with $n_0=0$, obtaining the effective Lagrangian $L_{eff}$:
\begin{align}
  \label{efflag0}
  L_{eff}=A_q^2\bigg[&-\frac{\partial \varphi}{\partial \xi}\coth\mu-\frac{1}{2}\frac{\partial b}{\partial \xi}\frac{1}{\sinh\mu}
\nonumber\\
&-2\coth\mu+2\frac{\cos b}{\sinh\mu}+A_q^2 N(\mu,\kappa)\bigg]
\end{align}
with non local contribution $N(\mu,\kappa)$:
\begin{align}
\label{fff}
N(\mu&,\kappa)=\frac{1}{A_q^4}\sum_{m,n}\frac{\exp(-\kappa\lvert m+n\rvert)}{\coth(\kappa/2)}\lvert q_m\rvert^2\lvert q_n\rvert^2=
\nonumber\\
&\frac{\tanh(\kappa/2)(2\sinh2\mu+\sinh\kappa+\sinh(4\mu+\kappa))}{4\sinh2\mu\sin^2(\mu+\kappa/2)}
\end{align}
as calculated with a bilateral Z-transform. Setting to zero each variational derivative of soliton parameters $(A_q,\varphi,b,\mu)$, we obtain the evolution of both soliton chirp $b$ and soliton width $\mu$. After some algebra:
\begin{align}
\label{lag0}
&\frac{\partial b}{\partial\xi}=\frac{4\cos b\sinh^3\mu}{\cosh2\mu}
\nonumber\\
&-W_0\frac{\sinh^2\mu\big[4N(\mu,\kappa)\tanh\mu+2\sinh^2\mu\frac{\partial N(\mu,\kappa)}{\partial\mu}\big]}{\cosh2\mu}\\
\label{lag1}
&\frac{\partial \mu}{\partial\xi}=-\frac{4\cosh\mu\sin b\sinh^2\mu}{\cosh(2\mu)}
\end{align}
with the soliton power $W_0=\sum_n\lvert q_n\rvert^2=A_q^2\coth\mu$. 
Equations (\ref{lag0})-(\ref{lag1}) define a two-dimensional phase-space with the conserved Hamiltonian $H_{eff}$:
\begin{widetext}
\begin{align}
  \label{effham0}
  H_{eff}=-W_0\bigg[-2+2\cos b&\mathrm{sech}\mu+W_0\frac{\tanh(\kappa/2)\tanh^2\mu(2\sinh2\mu+\sinh\kappa+\sinh(4\mu+\kappa))}{4\sinh2\mu\sin^2(\mu+\kappa/2)}\bigg]
\end{align}
\end{widetext}
Solitons correspond to stationary points of Eqs. (\ref{lag0})-(\ref{lag1}) with $b=0$ and:
\begin{align}
\label{ppp}
W_0&=\frac{8\cosh^2\mu\cosh\frac{k}{2}}{3\sinh^2\frac{k}{2}}\nonumber\\
&\times\frac{\sinh\mu\sinh^3(\mu+\frac{\kappa}{2})}{[\sinh\mu+\sinh(3\mu+\kappa)(-1+2\cosh2\mu)]}
\end{align}
Since $\frac{\partial W_0(\mu,\kappa)}{\partial \mu}>0$, all fixed points representing solitons are stable. \cite{wa_Malom}  As shown in Fig. \ref{power}a, the existence curve (\ref{ppp}) of discrete solitons rapidly approaches the local Kerr-case for diminishing non locality (i.e., increasing $\kappa>3$). As non locality is enhanced and $\kappa$ reduces towards and below 1, however, the refractive perturbation becomes broader and broader (in $y$) and $W_0$ larger and larger. Substantial changes are visible near the soliton solution, as displayed in the phase-plane of Eqs. (\ref{lag0})-(\ref{lag1}) in Fig. \ref{power}(b)-(d). In a Kerr regime ($\kappa\rightarrow\infty$) the phase-plane consists of a series of periodic orbits near the localized state $(\mu=1.5,b=0)$ and $\mu$ tends to zero for higher chirps b (Fig. \ref{power}b). Therefore, the addition of an initial chirp above a certain value -i.e., enough \emph{chirp imprinting}- destroys the soliton \cite{chirp_Kaup,chirp_DesaiA,wa_BEC_AbdulA}. The situation keeps unchanged as $\kappa\geq 3$ (Fig. \ref{power}c). In the non local regime ($\kappa=1.0$), conversely, the trajectories evolve from a closed-loop to a limit-cycle, hence no \emph{chirp-imprinting} can break the soliton (Fig \ref{power}d). This remarkable finding is confirmed by numerical simulations, as visible in Fig. \ref{sim1}. While a local system cannot sustain discrete light localization with an input spatial chirp above a threshold (Fig. \ref{sim1}a), non locality allows for the propagation of chirped discrete solitary waves (Fig. \ref{sim1}b) with periodically varying width, as predicted by our model. Clearly, the soliton amplitude oscillates as well in order to conserve the total power $W_0$. These novel solutions belong to the class of discrete breathers and are the lattice counterparts of the continuously breathing solitons reported in highly non local bulk NLC \cite{non_NLC_ContB}.\\
\begin{figure}
\includegraphics{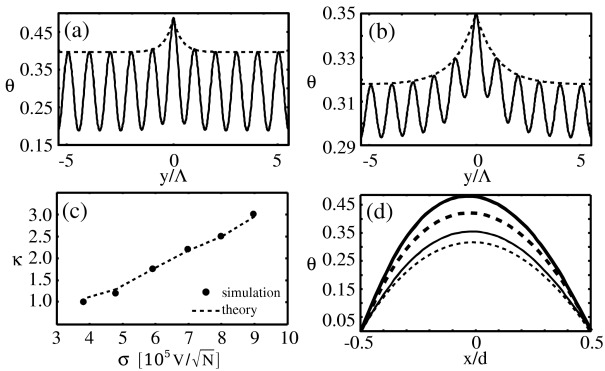}
\caption{\label{sim2}
Calculated reorientation distribution along $y$ ($x=0$) (solid line) and theoretical Green-function envelope (dashed line) in (a) the local ($\sigma=3.85\times 10^{5}\mathrm{V/\sqrt N}$) and in (b) the non local ($\sigma=9\times 10^{5}\mathrm{V/\sqrt N}$) cases, respectively; (c) corresponding degree of non locality $\kappa$ versus $\sigma$ after a fit(dots) and from theory (dashed line); (d) reorientation profile versus $x$ ($y=0$) without (dashed lines) and with (solid lines) optical excitation (1 mW) for local (thick lines) and non local responses (thin lines), respectively.} 
\end{figure}

\section{Numerical results with the NLC lattice}
\label{nres}
In order to link the analysis to an actual NLC lattice, we need to estimate the range of available $\kappa$. As it stems from the model above, the nonlinear index change has a peaked envelope of transverse size $\kappa=\frac{\Lambda}{R_c}\propto \Lambda V/\sqrt Kd$ (being $E_x(x)\approx V/d$ and $\theta_r\approx\mathrm{const}$) and, therefore, non locality can be tuned by acting on either one of the form-factor $\Lambda/d$, the bias $V$, the elastic constant $K$, the temperature \cite{NLC_PeccN}. With reference to a standard NLC (nematic 5CB, with $K=3.8\times 10^{-12}$N), to evaluate the Green function along $Y$ we employed an optical ($\lambda=1.064\mu$m) excitation $A(x,y)$ consisting of a Dirac distribution across $y$ and a Gaussian beam of waist $\approx d$ across $x$. We numerically integrated Eqs. (\ref{dir}) and (\ref{dirc}) using the relevant potential distribution (see Ref. \cite{wa_NLC_fratSA}), and finally derived the size $\kappa$ versus $\sigma \equiv \Lambda V/\sqrt Kd$ by fitting the calculated reorientation profile. As the material is tuned from local (Fig. \ref{sim2}a with $d/\Lambda=0.44$, $V=0.78$V) to non local (Fig. \ref{sim2}b with $d/\Lambda=1.0, V=0.75$V), the Green-function envelope of the nonlinear reorientation along y (solid line) is well approximated by our theoretical model (dashed line). In the latter analysis we employ both geometric $(d/\Lambda)$ and material $(V)$  tuning, slightly adjusting the bias to keep $\theta_r\approx 0.35$. This results in a quasi-linear transition (see Fig. \ref{sim2}c) from a local to a non local response as $\sigma$ varies. The corresponding transverse size of the non local response, represented by dots in Fig. \ref{sim2}c, shows that theory and numerics are in excellent agreement in the range covering local $(\kappa=3.0)$ to non local $(\kappa=1.0)$ responses. Moreover, the all-optical reorientation across $x$, vi
sible in Fig. \ref{sim2}d, is nearly sinusoidal (dashed line) and does not widen significantly when the nonlinearity intervenes (solid line), supporting the validity of the local approximation previously adopted.\\

\section{Conclusions}
In conclusion, for the first time to the best of our knowledge and with specific reference to nematic liquid crystals and their reorientational response, we have modeled discrete light localization in a nonlinear medium with an arbitrary degree of transverse non locality. Starting from the governing equation of the liquid crystalline system, we performed an original reduction to a novel general form of discrete nonlinear non local Schr\"odinger equation. Remarkably, the latter result was not achieved by introducing an \emph{a priori} specific from of non locality, \cite{non_LAT_GaidiA,non_LAT_KartaA,non_DIS_GaidiA,non_DIS_Fedd} but one derived from the molecular response of NLC. We employed a variational procedure and investigated the role of non locality in supporting chirp-imprinted discrete spatial solitons. Such novel solutions are periodic breathers and cannot exist in purely local systems. Since the degree of non locality in NLC arrays can be adjusted by acting on geometric or material or external parameters \cite{NLC_PeccN}, we anticipate that our findings will trigger the observation of discrete light propagation in both local and non local regimes in one and the same system. Our numerical experiments, in excellent agreement with the theoretical predictions, fully support such possibility.\\
We acknowledge enlightening discussions with C. Conti and D. Levi.


\end{document}